\documentclass[%
 aip,
rsi,%
preprint,%
]{revtex4-1}

\usepackage{graphicx}% Include figure files
\usepackage{subfigure}
\usepackage{dcolumn}% Align table columns on decimal point
\usepackage{tabularx}
\usepackage{caption}
\usepackage{multirow}

\begin{document}

%\preprint{AIP/123-QED}

\title{An integrated device with high performance multi-function generators and Time-to-digital convertors}% Force line breaks with \\
%\thanks{Footnote to title of article.}

\author{X. Qin}
 \altaffiliation{X. Qin and Z. Shi contributed equally to this work.}
 \affiliation{ 
CAS Key Laboratory of Microscale Magnetic Resonance and Department of Modern Physics, University of Science and Technology of China, Hefei 230026, China.
}
 \affiliation{ 
Synergetic Innovation Center of Quantum Information and Quantum Physics, University of Science and Technology of China, Hefei, 230026, China.
}

\author{Z. Shi}
 \altaffiliation{X. Qin and Z. Shi contributed equally to this work.}
 \affiliation{ 
CAS Key Laboratory of Microscale Magnetic Resonance and Department of Modern Physics, University of Science and Technology of China, Hefei 230026, China.
}

\author{Y. Xie}
 \affiliation{ 
CAS Key Laboratory of Microscale Magnetic Resonance and Department of Modern Physics, University of Science and Technology of China, Hefei 230026, China.
}
\author{L. Wang}
 \affiliation{ 
CAS Key Laboratory of Microscale Magnetic Resonance and Department of Modern Physics, University of Science and Technology of China, Hefei 230026, China.
}

\author{X. Rong}%
 \email{xrong@ustc.edu.cn, djf@ustc.edu.cn}
 \affiliation{ 
Synergetic Innovation Center of Quantum Information and Quantum Physics, University of Science and Technology of China, Hefei, 230026, China.
}

\author{W. Jia}
 \affiliation{ 
CAS Key Laboratory of Microscale Magnetic Resonance and Department of Modern Physics, University of Science and Technology of China, Hefei 230026, China.
}
\author{W. Zhang}
 \affiliation{ 
CAS Key Laboratory of Microscale Magnetic Resonance and Department of Modern Physics, University of Science and Technology of China, Hefei 230026, China.
}

\author{J. Du}
 \email{xrong@ustc.edu.cn, djf@ustc.edu.cn}
\affiliation{ 
CAS Key Laboratory of Microscale Magnetic Resonance and Department of Modern Physics, University of Science and Technology of China, Hefei 230026, China.
}
\affiliation{ 
Synergetic Innovation Center of Quantum Information and Quantum Physics, University of Science and Technology of China, Hefei, 230026, China.
}

\date{\today}% It is always \today, today,

\begin{abstract}
A highly integrated, high performance, and re-configurable device, which is designed for the Nitrogen-Vacancy center based quantum applications, is reported. The digital compartment of the device is fully implemented in a Field-Programmable-Gate-Array. The digital compartment is designed to manage the multi-function digital waveform generation and the Time-to-Digital-Convertors. The device provides two Arbitrary-Waveform-Generator channels which operate at a 1 Gsps sampling rate with a maximum bandwidth of 500 MHz. There are twelve pulse channels integrated in the device with a 50 ps time resolution in both duration and delay. The pulse channels operate with the 3.3 V Transistor-Transistor logic. The FPGA-based Time-to-Digital-Convertor provides a 23-ps time measurement precision. A data accumulation module, which can record the input count rate and the distributions of the time measurement, is also available. A Digital-to-Analog-Convertor board is implemented as the analog compartment, which converts the digital waveforms to analog signals with 500 MHz low-pass-filters.  All the input and output channels of the device are equipped with 50 Ω Sub-Miniature version A termination. The hardware design is modularized thus it can be easily upgraded with compatible components. The device is suitable to be applied in the quantum technologies based on the N-V centers, as well as in other quantum solid state systems, such as quantum dots, phosphorus doped in silicon and defect spins in silicon carbide.
\end{abstract}

\maketitle

%\begin{quotation}

%\end{quotation}

\section{MOTIVATION}

Development of the semi-conductor technology allows the manufacture of high capacity Field-Programmable-Gate-Arrays (FPGA), which are Application-Specific-Integrated-Circuits (ASIC) to be configured by customers. The re-programmable FPGA chips provide opportunities to implement high flexible, highly integrated and highly synchronized devices for scientific researches. Such applications include quantum computation \cite{bernien2013heralded,maurer2012room,dutt2007quantum,neumann2008multipartite,robledo2011high,waldherr2014quantum,weber2010quantum,wrachtrup2010defect} and quantum metrology  \cite{balasubramanian2008nanoscale,maze2008nanoscale,taylor2008high,dolde2011electric,neumann2013high,toyli2013fluorescence,kucsko2013nanometre,shi2014sensing,shi2015single} based on Nitrogen-Vacancy (N-V) center in diamonds. High performance multi-channel Arbitrary-Waveform-Generators (AWG) and pulse generators are usually required to realize precise quantum control \cite{rong2015experimental,goldman2015phonon} and Time-to-Digital-Convertors (TDC) are used to detect the quantum states of the N-V center \cite{robledo2010control,qin2016high}. The existing solution is to use independent components to implement the multi-function device. Such method contributes to an ``open-loop'' system, which is very complex, expensive and lacking of synchronism. On the other hand, the coherent time of the electron spin in the N-V center is very short, ranging from microseconds to milliseconds \cite{balasubramanian2008nanoscale,maze2008nanoscale,taylor2008high}. The long latency and the low synchronization of the ``open-loop'' system makes it very difficult to achieve some real-time feedback operations. All these disadvantages restrict further developments of quantum technologies based on the N-V centers. Therefore, a high performance ``close-loop'' device which is high flexible and highly integrated, is needed to overcome the restrictions in the future development of the related quantum applications.

The FPGA allows a possible solution to develop powerful devices with high flexibility and integration. In this paper, we present an integrated “close-loop” device utilizing implemented in the FPGA for applications based on N-V centers. The AWGs, the high time resolution pulse generators and the high precision TDCs are fully implemented using a single Xilinx Virtex-7 FPGA, which is installed in a 12-layer printed-circuit-board (PCB).

\section{HARDWARE}

\begin{figure}
  \center{\includegraphics[width=\textwidth]{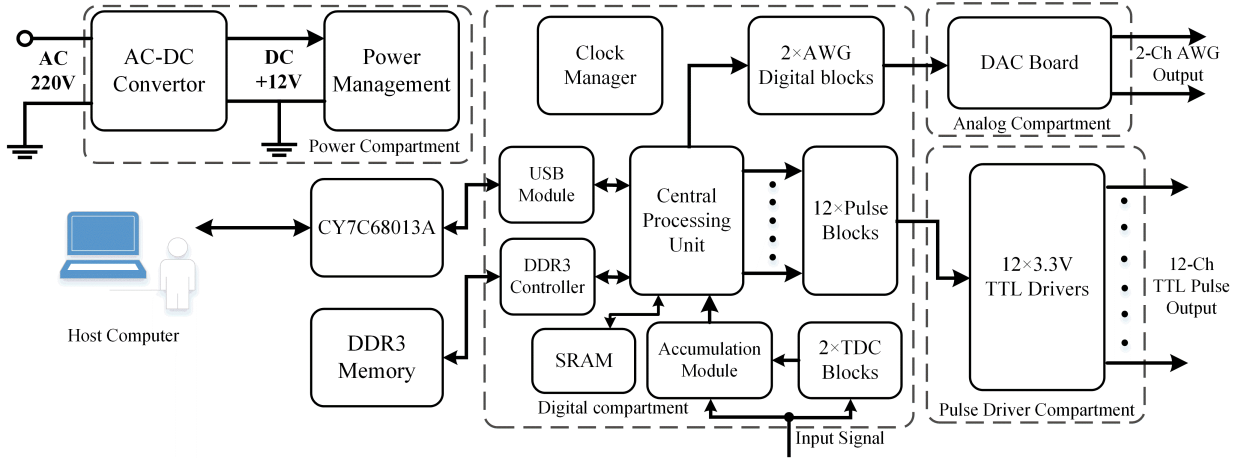}}
  \caption{The block diagram of the integrated device for the N-V center based quantum applications.}
  \label{Fig.1}
\end{figure}

The block diagram of the integrated device is shown in Figure \ref{Fig.1}. The digital compartment is implemented using an FPGA, which performs the central management of the digital waveform generation and the Time-to-digital conversion. The pulse blocks can continuously generate digital pulse signals with a 50 ps time resolution. The pulse generators are equipped with 3.3V Transistor-Transistor logic (TTL) drivers, which are installed in the pulse driver compartment. The digital arbitrary waveform data are pre-stored in an external Double-data-rate (DDR) Memory with 1 Giga-Bytes (GB) capacity. A DDR3 controller is implemented in the FPGA to manage the memory. The AWG digital block manages the waveform generation and outputs the parallel 16-bit digital signals with a 1 Gsps sampling rate. The photons generated from N-V center are fed to an Avalanche-Photon-Diode and then photon events can be measured by the TDC blocks. The TDC measures the arrival time of photons with a 23 ps precision, and provides an accumulation module which records the count rate and the distribution of the arrival time. The pulse data and the TDC data are stored in the SRAM integrated in the FPGA. The central processing unit manages the command distribution and the data transmission for the digital compartment. The clock manager generates the system clocks for the FPGA logics. The host computer communicates with the central control block via a Universal-Serial-Bus (USB) controller CY7C68013A. The USB module in the FPGA manages the communication with the USB controller. The whole device is powered by an off-the-shelf 12 V power supply, and all the power supplies are located in the power compartment. The analog compartment is a 16-bit Digital-to-Analog (DAC) board, which outputs the analog waveforms with 500 MHz Low-Pass-Filters (LPF). The power rails for the Sub-Miniature version A (SMA) digital circuits are generated by switching voltage regulators, and the analog circuits are powered by the Low-Drop-Out (LDO) regulators, equipped with pi type filters. The integration of signal generation and event measurement in a single device contributes to a ``close-loop'' system when used in the quantum applications based on N-V centers. The overall latency between sending the excitation pulse and receiving the response signal is approximately hundreds of nanoseconds. Unnecessary cable length will lead to extra latency, approximately 0.25 ns per inch. On the other hand, all the modules operate with the same pattern thus the system has a precise timing in synchronism.

\begin{figure}
  \center{\includegraphics[width=\textwidth]{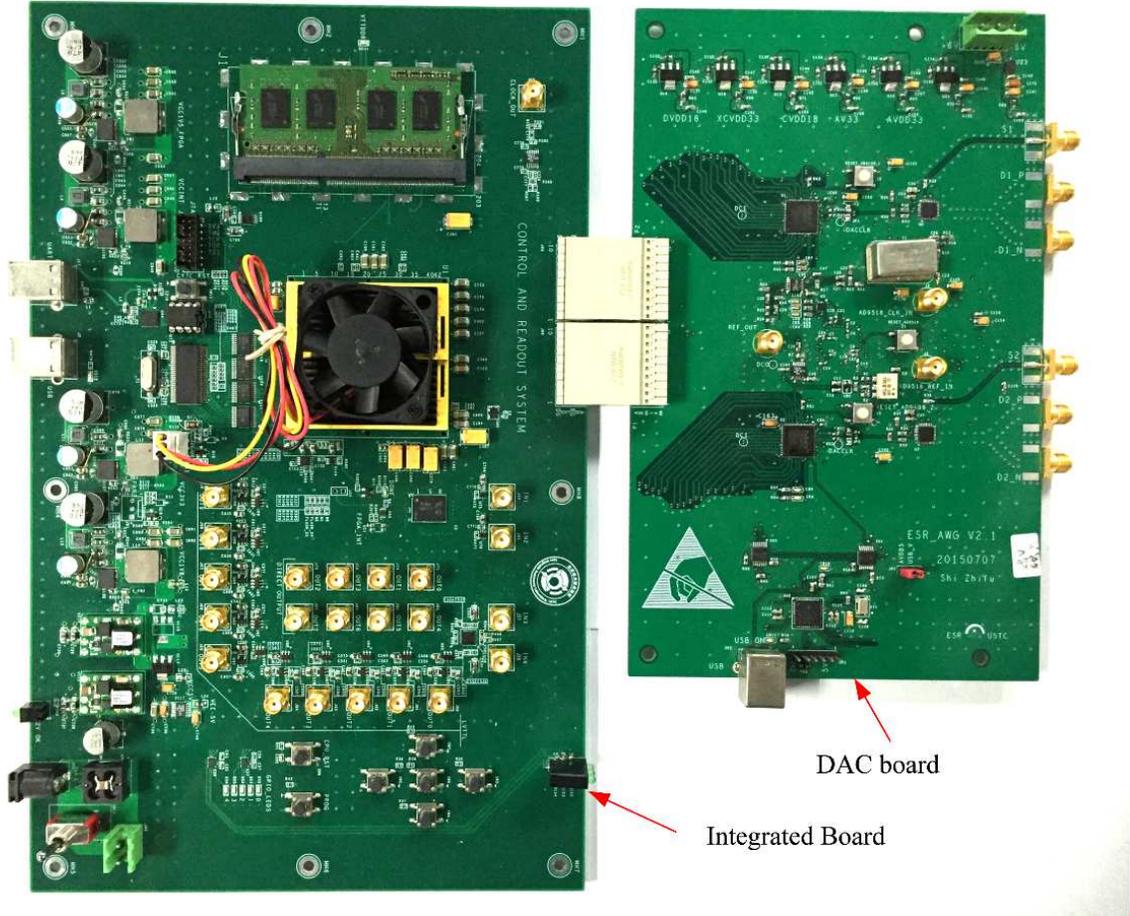}}
  \caption{Photograph of the integrated device. The PCB on the left is the integrated board which manages the digital control of the signal generation and readout; The PCB on the right side is the DAC board which can convert the digital waveforms to analog signals.}
  \label{Fig.2}
\end{figure}

The digital compartment, power compartment, pulse driver compartment, USB controller and DDR3 memory of the integrated device are assembled on a 12-layer integrated PCB, while the DAC board is a 10-layer PCB. Figure \ref{Fig.2} shows the photograph of this device. The PCB on the left is the integrated board, and the DAC board is on the right side. A Xilinx Virtex-7 FPGA (XC7VX485T-ffg1761) is used for the digital control of the signal generation and readout. These two boards are connected by two AMP-1469169-1 connectors. All the input and output channels of the two boards are equipped with 50 $\Omega$ SMA terminations. The hardware design of the device is modularized, thus it is easy to upgrade the device by re-configuring the digital compartment, or replacing the FPGA chip and the DAC board which are designed with the same compatible interface specifications.

\subsection{Arbitrary waveform generator}

\begin{figure}
  \center{\includegraphics[width=\textwidth]{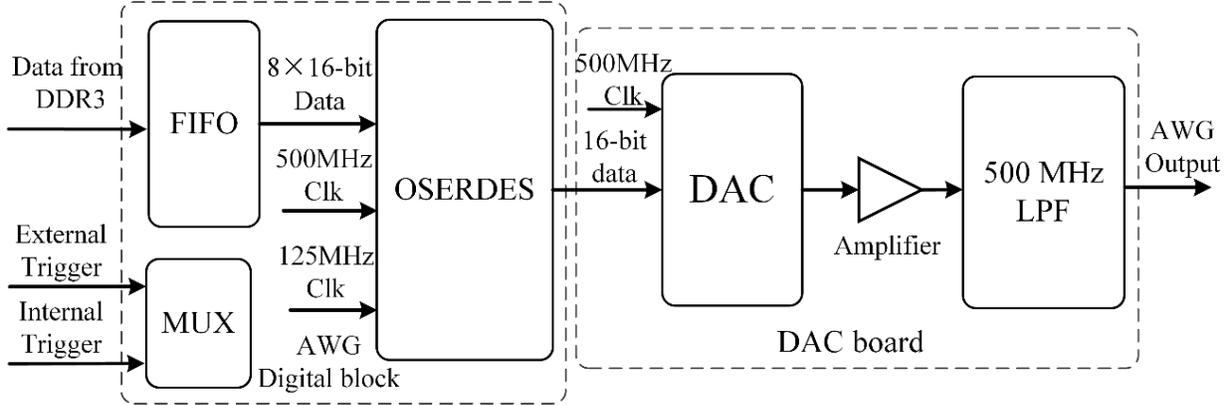}}
  \caption{The schematic diagram of the arbitrary waveform generator.}
  \label{Fig.3}
\end{figure}

The schematic diagram of the FPGA-based arbitrary waveform generator is shown in Figure \ref{Fig.3}. The AWG data are pre-stored in the external DDR3 memory. The AWG digital block loads the waveform data from the DDR3 memory and cache them in the First-In-First-Out (FIFO) memory in the FPGA. At each rising edge of the 125 MHz clock, the OSERDES module loads 8$\times$16-bit waveform data, and outputs parallel 16-bit data at each edge of the 500 MHz clock with the Double-Data-rate mode, which contributes to a 1 Gsps sampling rate. The interface between the FPGA and the DAC board is designed with the Low-Voltage-Differential-Signaling (LVDS) standard. There are 34 differential pairs in the interface, 32 of them are digital signals of the 2-channel AWG, and the other two are 500 MHz sampling clocks for the DAC. Two AD9139 ASIC chips are implemented in the DAC board for digital-to-analog conversion with a 16-bit amplitude resolution and an analog bandwidth up-to 500 MHz. The analog circuits in the DAC board can amplify and filter the output signals, with a 2 V peak-to-peak output voltage. The corresponding noise spectral density of the amplifiers is -161.5 dBm/Hz, and the 11-order low pass Butterworth-filter provides a 500 MHz cut-off frequency. The AWG can operate with a switchable trigger mode and two trigger signals are available for each channel. The external trigger is an off-board signal which is used as an asynchronous trigger event. The internal trigger is produced by the FPGA logics and it can be used in synchronous waveform generation.

\subsection{Pulse generator}

\begin{figure}[h]
  \center{\includegraphics[width=\textwidth]{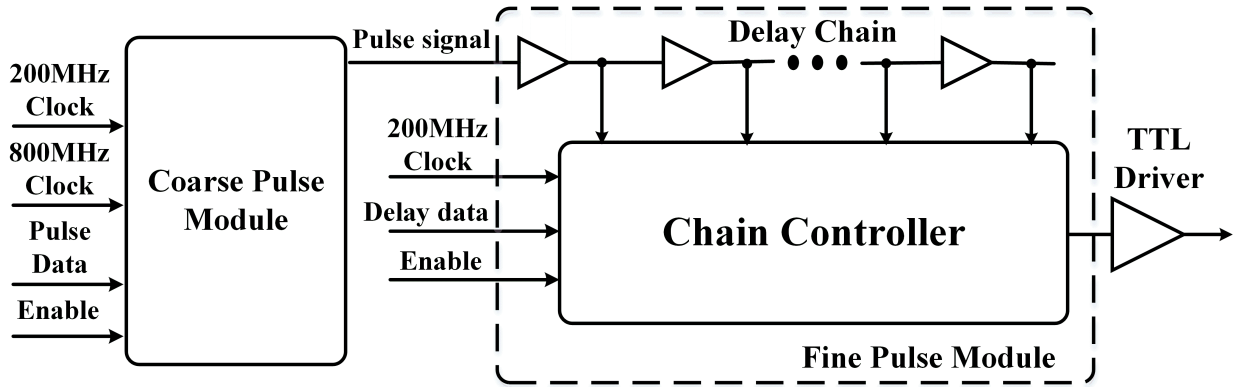}}
  \caption{ The basic diagram of the high time resolution pulse generator.}
  \label{Fig.4}
\end{figure}

The pulse generator in the device is based on time interpolating method \cite{goldman2015phonon}. A delay chain is implemented in the pulse channel for fine time interpolating. Figure \ref{Fig.4} shows the basic diagram of the pulse generator, the coarse pulse module and the fine pulse module are both integrated in the virtex-7 FPGA. The 200 MHz clock is applied in data processing, and an 800 MHz clock is used to generate high speed pulse outputs. The 80-bit pulse data consists of a double 32-bit data for the duration time of the `0' logic and the `1' logic, and two 8-bit data for the fine delay time of the rising and falling edges. The coarse pulse module processes the pulse data and outputs pulse signals with a 1.25 ns time resolution, which equals the period of the 800 MHz clock. The pulse signals will go through the delay chain in the fine pulse module before being output, and the output position can be real-time adjusted at each leading edge of the 200 MHz clock. The averaged delay time of the delay cells is around 50 ps, and the total delay of the chain should cover the period of the 800 MHz clock. Such solution contributes to a non-dead-time pulse generator, which is characterized by a 50 ps resolution, a 5 ns minimum pulse width, and a dynamic range exceeding two seconds. The single-bit transceiver SN74AVC1T45 and the high-speed buffer BUF602 are applied as the TTL driver circuits. The single-bit transceiver is used to covert the 1.8 V LVCOMS signal from the FPGA to 3.3 V TTL electrical level, and BUF602 is applied to provide the capacity for driving the 50 ohm terminal resistance, with rising and falling edges shorter than 1 ns.

\subsection{Time-to-digital convertor}

\begin{figure}[h]
  \center{\includegraphics[width=\textwidth]{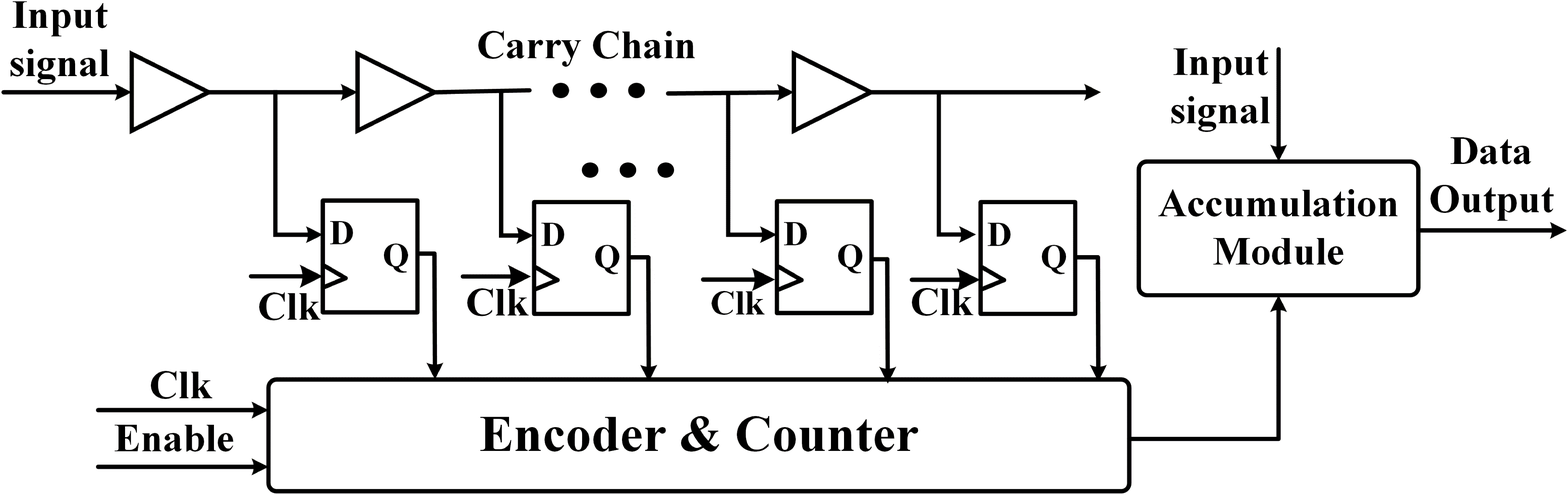}}
  \caption{Block diagram of the FPGA-based TDC.}
  \label{Fig.5}
\end{figure}

TDCs are used to record the arriving time of the output photon from N-V centers. The FPGA-based high time resolution TDC is implemented via a cascaded carry chain \cite{song2006high,wang2010fully,qin2013development} in the virtex-7 FPGA, and Figure \ref{Fig.5} shows the block diagram of the TDC. The TDC block operates with a 125 MHz system clock. A counter is used to record the coarse time of the input signal, and the carry chain is used for fine time measurement. Each input signal will transmit through the carry chain. The exact position of the signal leading edge will be recorded by the D-flip-flop (DFF) group. The DFF group outputs a thermometer code after measuring the arriving time of the input signal. The thermometer code is a unary code which is composed of several ones followed by zeros, and the amount of the ones stands for the precise time interval between the clock rising edge and the signal arrival moment. The encoder is used to convert the thermometer code to binary code and generate the fine time data. The coarse time data and the fine time data are both managed by the accumulation module before output to PC. The frequency of the operating clock for the TDC module is 125 MHz, thus the total delay of the carry chain should cover the 8 ns clock period. This TDC can achieve a time resolution of 23 ps with a dynamic range of 42 seconds. A high speed comparator LMH7322, which contributes to a 170 ps fast rise time, is used to translate the input signal into the LVDS electrical level before measured by the TDC. The accumulation module is integrated for data acquisition, which can either record the counting rates of the input signals or record the distribution of the signal arriving time.

\subsection{Resource occupation}

\begin{table}[h]
  \small
  \newcommand{\tabincell}[2]{\begin{tabular}{@{}#1@{}}#2\end{tabular}}
  \newcolumntype{Y}{>{\centering\arraybackslash}X}

  \makebox[\textwidth][c]{
  \begin{tabularx}{\textwidth}[b]{Y|cYcYYYYY}
    \hline
    \hline
    Resource\textsuperscript{\emph{a}} & AWG  & \tabincell{c}{Pulse \\ Generator} & TDC & \tabincell{c}{DDR3 \\ Controller} & \tabincell{c}{USB \\ Controller} & \tabincell{c}{Total \\ Utilization} & Available\textsuperscript{\emph{b}} & Occupation  \\
    \hline
    6-input LUTs & \multirow{2}*{268} & \multirow{2}*{1493} & \multirow{2}*{2249} & \multirow{2}*{10847} & \multirow{2}*{67} & \multirow{2}*{31543} & \multirow{2}*{303600} & \multirow{2}*{10.39\%}  \\
    %\hline
    Slices       & 137 & 600  & 846  & 4873  & 43 & 13189 & 75900  & 17.38\%  \\
    %\hline
    Flip Flops   & 387 & 1623 & 2547 & 13309 & 88 & 36109 & 303600 & 11.89\%  \\
    %\hline
    I/Os         & 18  & 1    & 2    & 114   & 29 & 177   & 700    & 25.29\%  \\
    %\hline
    BRAMs (Kb)    & \multirow{2}*{4}   & \multirow{2}*{55}   & \multirow{2}*{15}   & \multirow{2}*{20}    & \multirow{2}*{1}  & \multirow{2}*{707}   & \multirow{2}*{1030}   & \multirow{2}*{68.64\%}  \\
    \hline
    \hline
  \end{tabularx}
  }
  \caption{Resource occupation of the integrated device. }
  \label{Tab.1}
  \begin{flushleft}
  \textsuperscript{\emph{a}} 6-input look-up-tables (LUT), slices and flip flops are the basic logic resources of the FPGA, the I/O ports are the user defined digital data channels. The block RAM (BRAM) memories are the internal storage cells in the FPGA;

  \textsuperscript{\emph{b}} Total amount of the available resources in the Virtex-7 FPGA XC7VX485T-ffg1761.
  \end{flushleft}
\end{table}

As all the signal generation and read-out functions of the integrated device are fully implemented using the FPGA resources, the resource occupation in the Virtex-7 FPGA XC7VX485T-ffg1761 is significant for design consideration. Table I shows the resource occupation of each module per single channel in the device, the resource utilization of the entire design, and the total amount of the available resources in the FPGA. 6-input Look-up-tables (LUTs), Slices, and flip flops are the basic logic resources of the Xilinx Virtex-7 FPGA. All the basic resources are re-programmable and can provide favorable flexibility to satisfy various requirements. The input and output (I/O) ports are the user defined digital data channels, and the block RAM memories are the internal storage cells integrated in the FPGA. The single channel resource occupation is the important reference when immigrating the modules into FPGAs with different specifications. On the other hand, there are plenty of unused FPGA logic resources remaining after completing the device implementation. Thus the integrated device gives a great possibility to realize the real-time digital “feedback” in the future, just by utilizing the unused FPGA resources and uploading the FPGA configuration, without any hardware modification.

\section{SOFTWARE}

\begin{figure}[h]
  \center{\includegraphics[width=\textwidth]{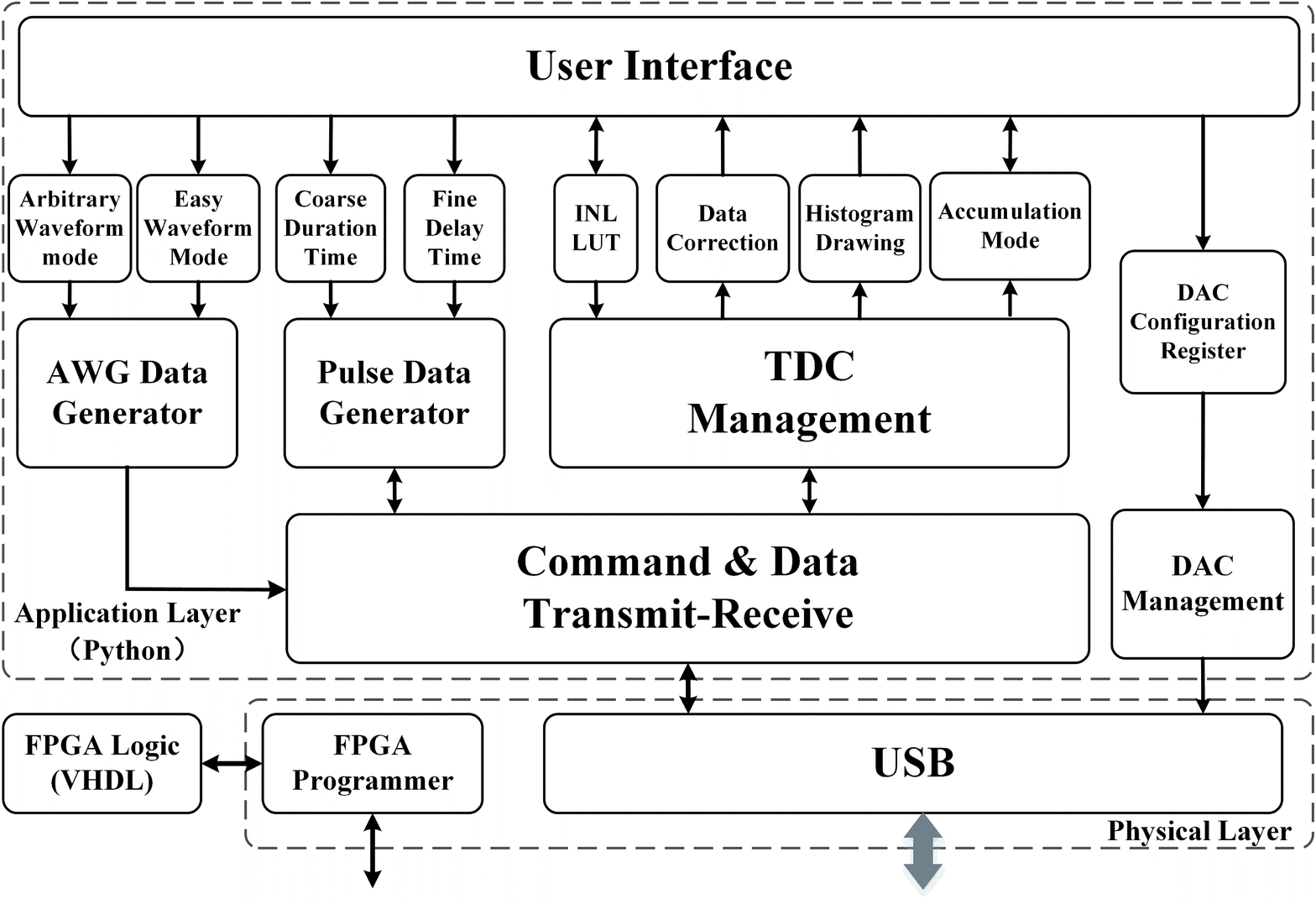}}
  \caption{The software architecture of the integrated device for N-V center based applications. The software manages the generation of the AWG data and Pulse data, the control of the TDC channels, and the DAC configuration.}
  \label{Fig.6}
\end{figure}

A customized software to handle the device has been developed. Figure \ref{Fig.6} shows the software architecture. The generation of the AWG data and Pulse data, the control of the TDC channels and the DAC configuration are fully managed by this software. Any arbitrary waveform data can be applied to the AWG with the `arbitrary waveform' mode. The `easy waveform' mode can be used to quickly generate some common waveforms such as Pulse, Triangular wave, Sine and Cosine wave, Gaussian wave, Zigzag wave and etc. The software can also guide the users to generate the digital data for the high resolution pulse channels. The channel calibration, the data correction, the histogram drawing and the data accumulation procession for the TDC channels is also managed by the software. Once the host PC is connected to the device via the USB bus, the software allows automatically configuration for the status registers of the DAC chips, and performing any further command.

\section{PERFORMANCE}

\subsection{Arbitrary waveform generator}

\begin{figure}[h]
  \centering
  \subfigure{
    \label{Fig.7a}
    \includegraphics[width=3.0in]{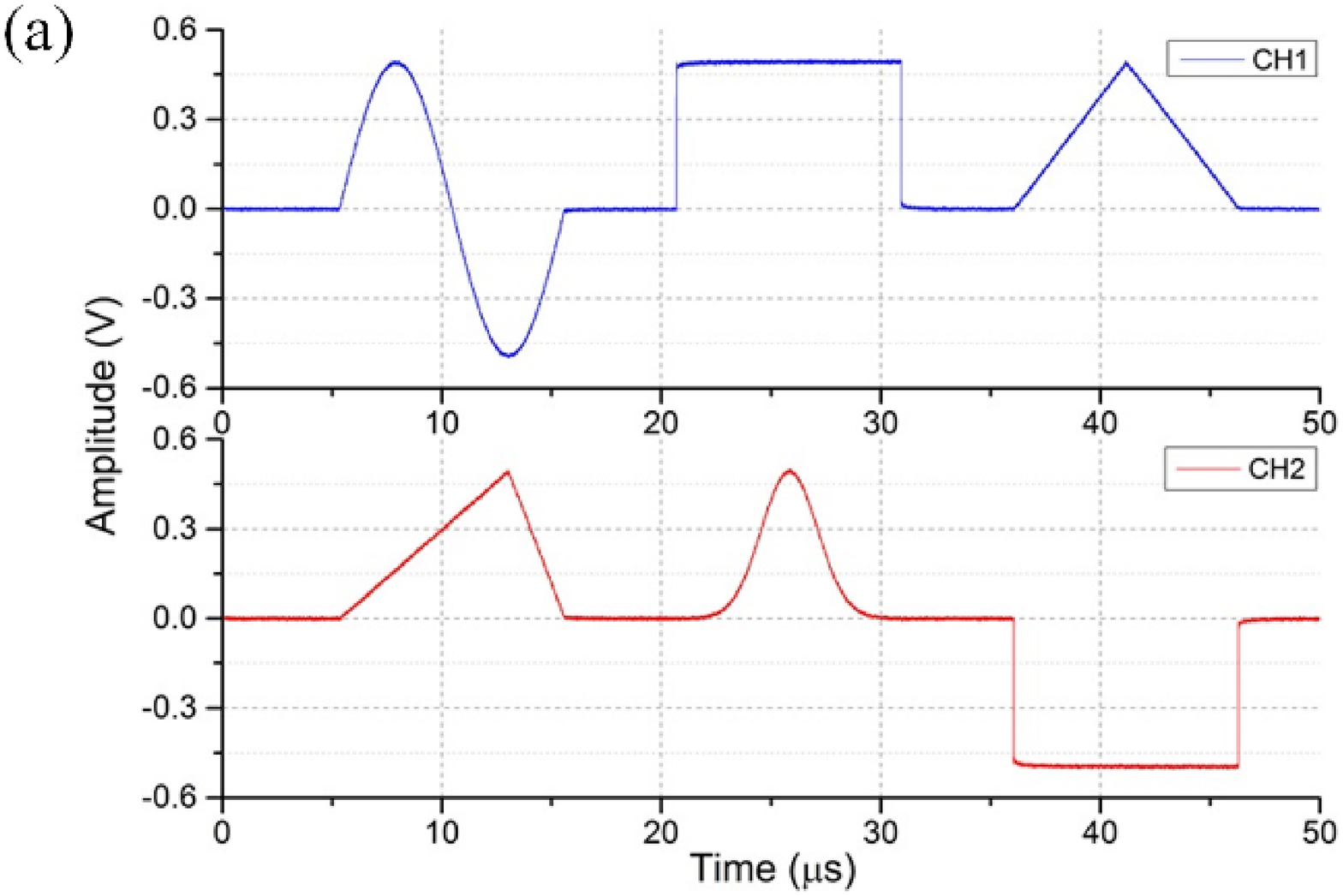}}
  %\hspace{1in}
  \subfigure{
    \label{Fig.7b}
    \includegraphics[width=3.0in]{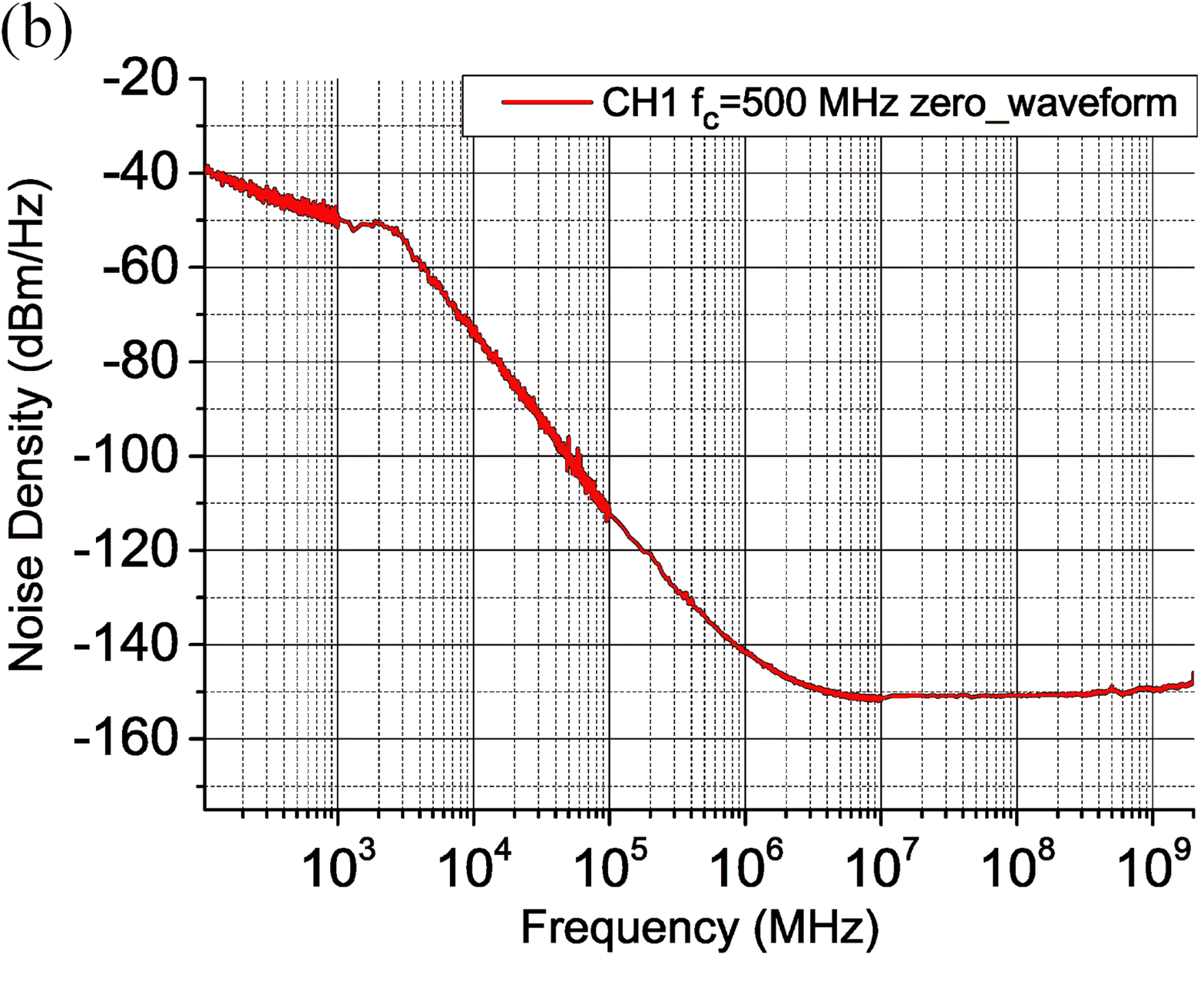}}
  \subfigure{
    \label{Fig.7c}
    \includegraphics[width=3.0in]{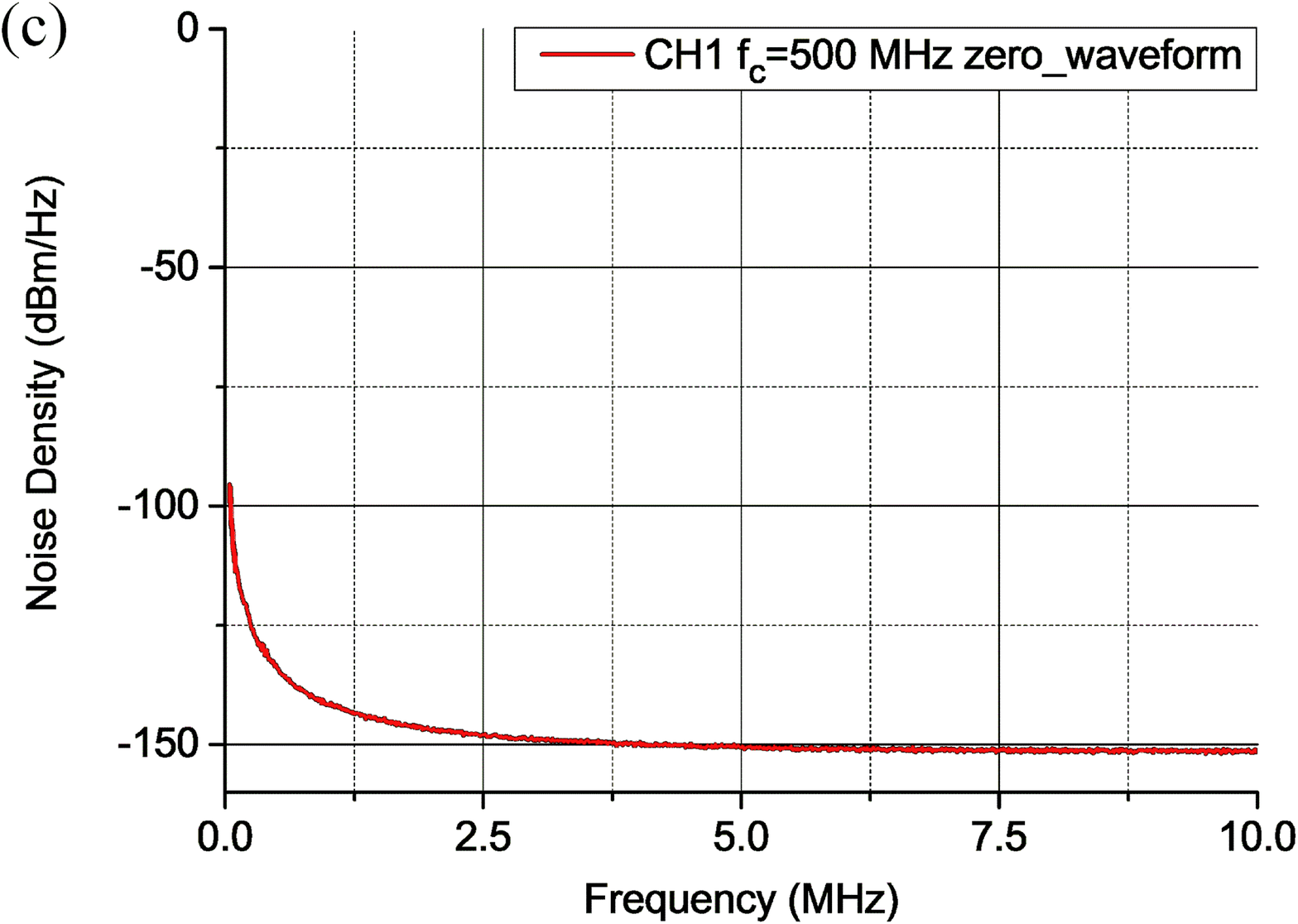}}
  \subfigure{
    \label{Fig.7d}
    \includegraphics[width=3.0in]{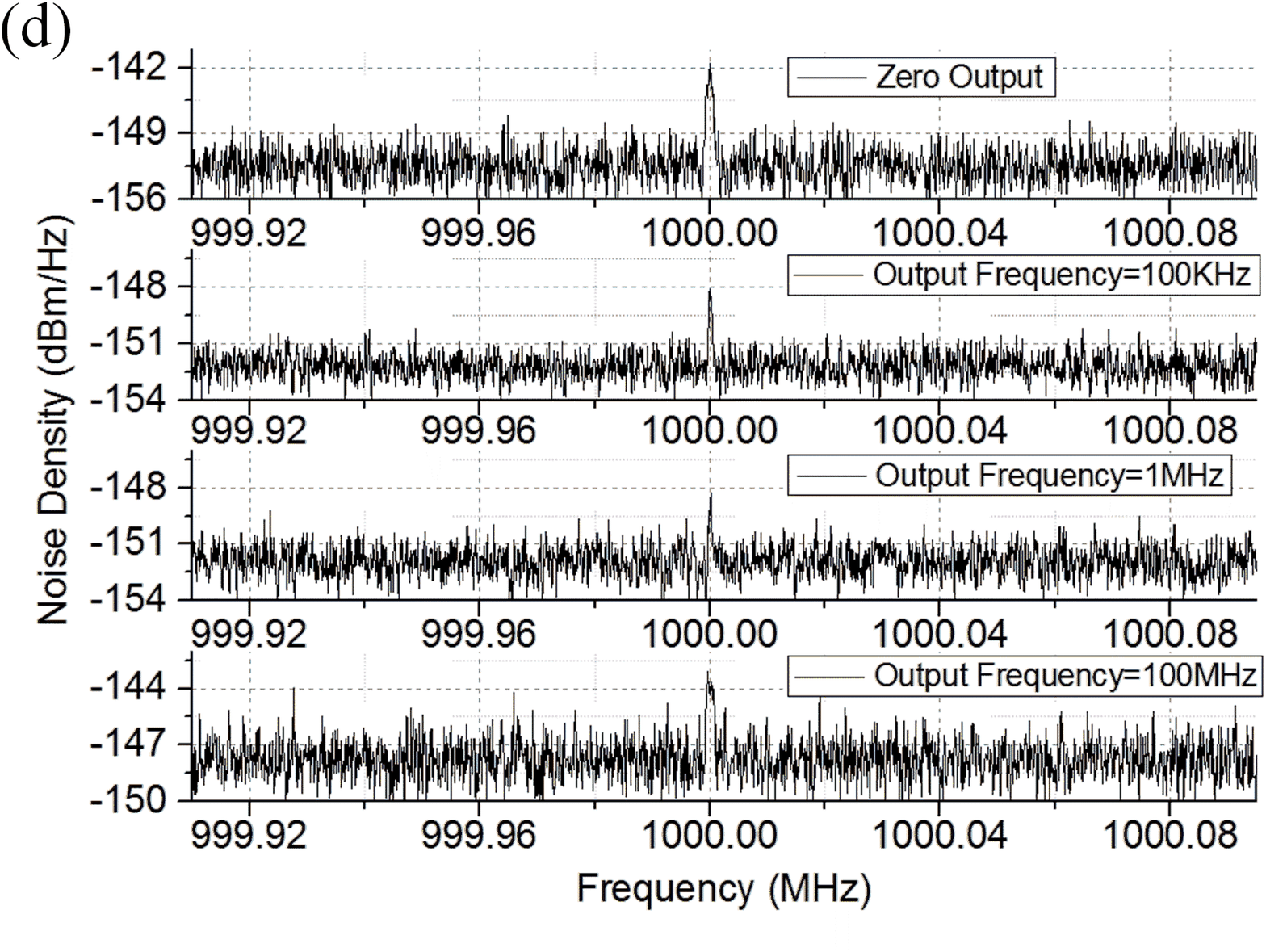}}
  \caption{Signal characteristic of the arbitrary waveform generator.(a) The self-defined waveform from two AWG channels. (b) The wide-band noise spectra over 500 MHz low pass filter measured up-to 2 GHz. (c) The low frequency noise spectra within 10 MHz. (d) The 1GHz clock noise when outputting zero waveform, 100 kHz, 1 MHz, and 100 MHz sinusoidal signals.}
  \label{Fig.7}
\end{figure}

The signal characteristics of the arbitrary waveform signals are demonstrated in Figure \ref{Fig.7}. A LeCroy WavePro 735 Zi digital oscilloscope with a 40 GS/s real-time sampling rate and a 3.5 GHz bandwidth was used to record the waveform. The waveform signals were generated using the `easy waveform' mode. Figure \ref{Fig.7a} shows the self-defined waveforms including sine wave, square wave, triangle wave, saw-tooth wave and Gaussian wave. The two AWG channels output the waveforms synchronously. The 1 GHz sampling clock noise and the low frequency noise from the power supply circuitry are the main noise contributors of the AWG. A Keysight N9020A Spectrum Analyzer was used to measure the noise spectra. As shown in Figure \ref{Fig.7b}, the noise spectra at channel one of the AWG was measured up-to a frequency of 2 GHz. Figure \ref{Fig.7c} demonstrates the low frequency noise power. The AWG was configured to output zero-waveform, and the noise density from the DC-DC convertors was suppressed to -150 dBm/Hz. Figure \ref{Fig.7d} shows the suppression of the 1GHz clock noise when outputting zero waveform, 100kHz, 1 MHz, and 100 MHz sinusoidal signals, and the noise density is less than -143 dBm/Hz at 1 GHz.

\subsection{Pulse generator}

\begin{figure}[hbp]
  \centering
  \subfigure{
    \label{Fig.8a}
    \includegraphics[width=6.0in]{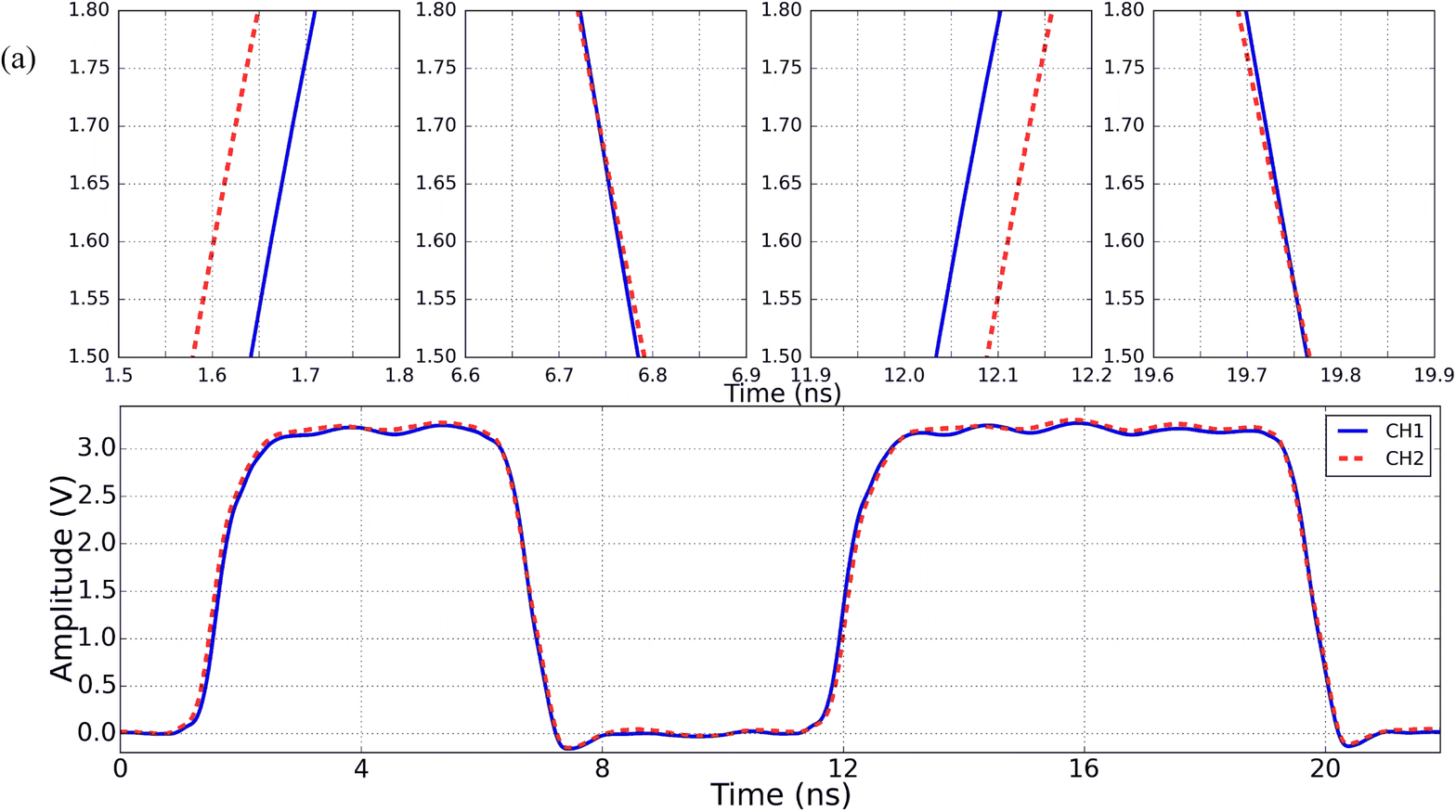}}
  %\hspace{1in}
  \subfigure{
    \label{Fig.8b}
    \includegraphics[width=3.0in]{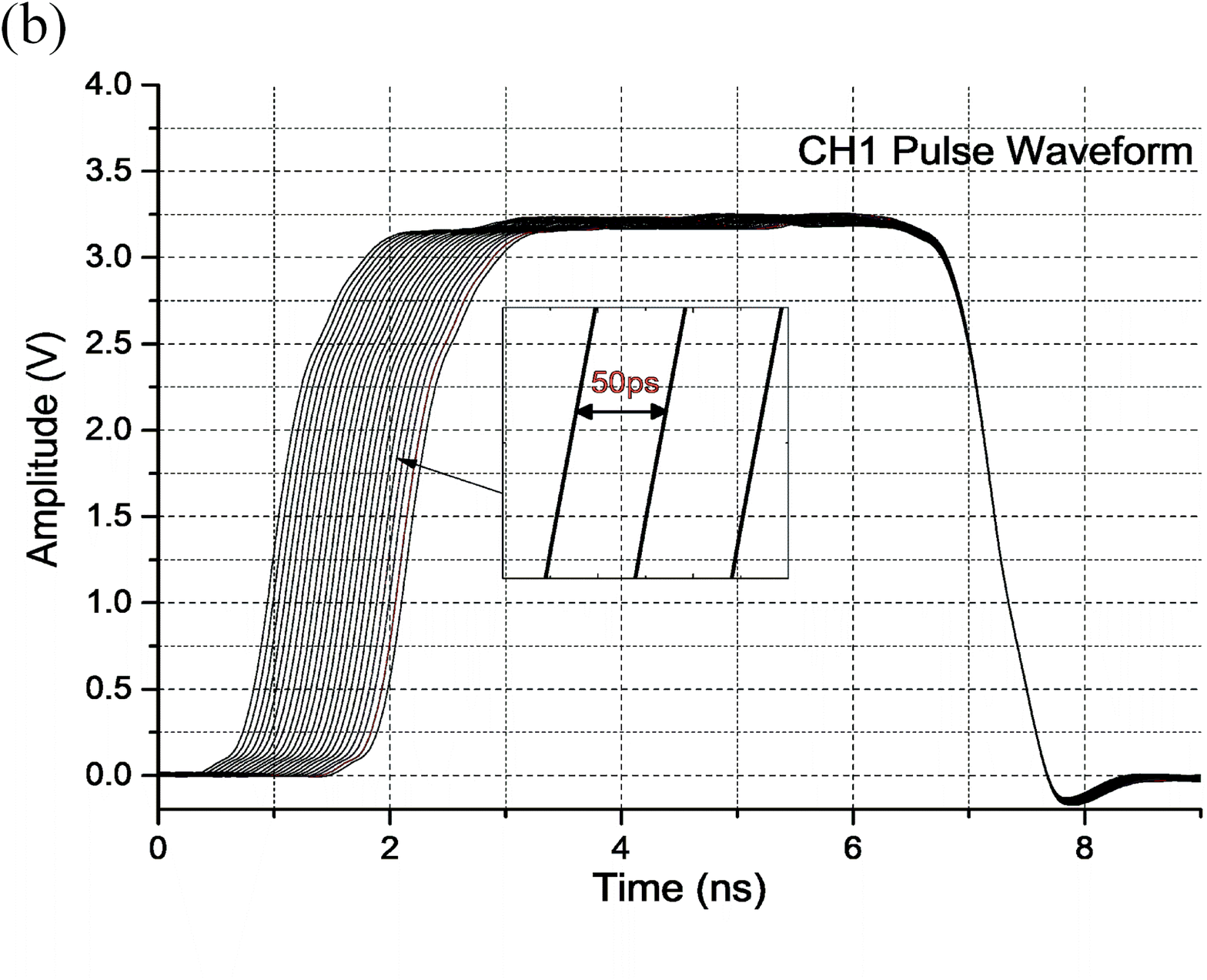}}
  \subfigure{
    \label{Fig.8c}
    \includegraphics[width=3.0in]{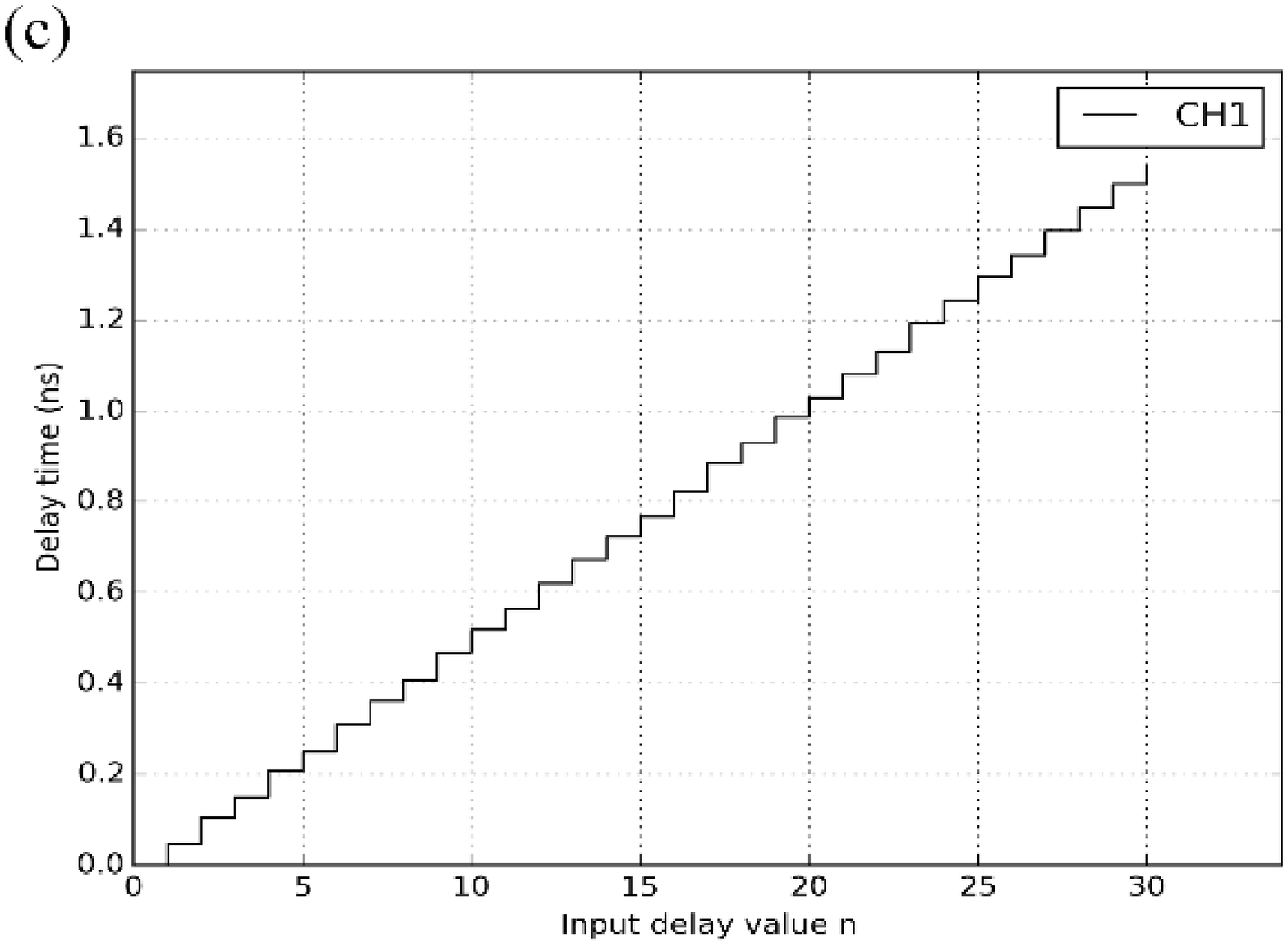}}
  \subfigure{
    \label{Fig.8d}
    \includegraphics[width=3.0in]{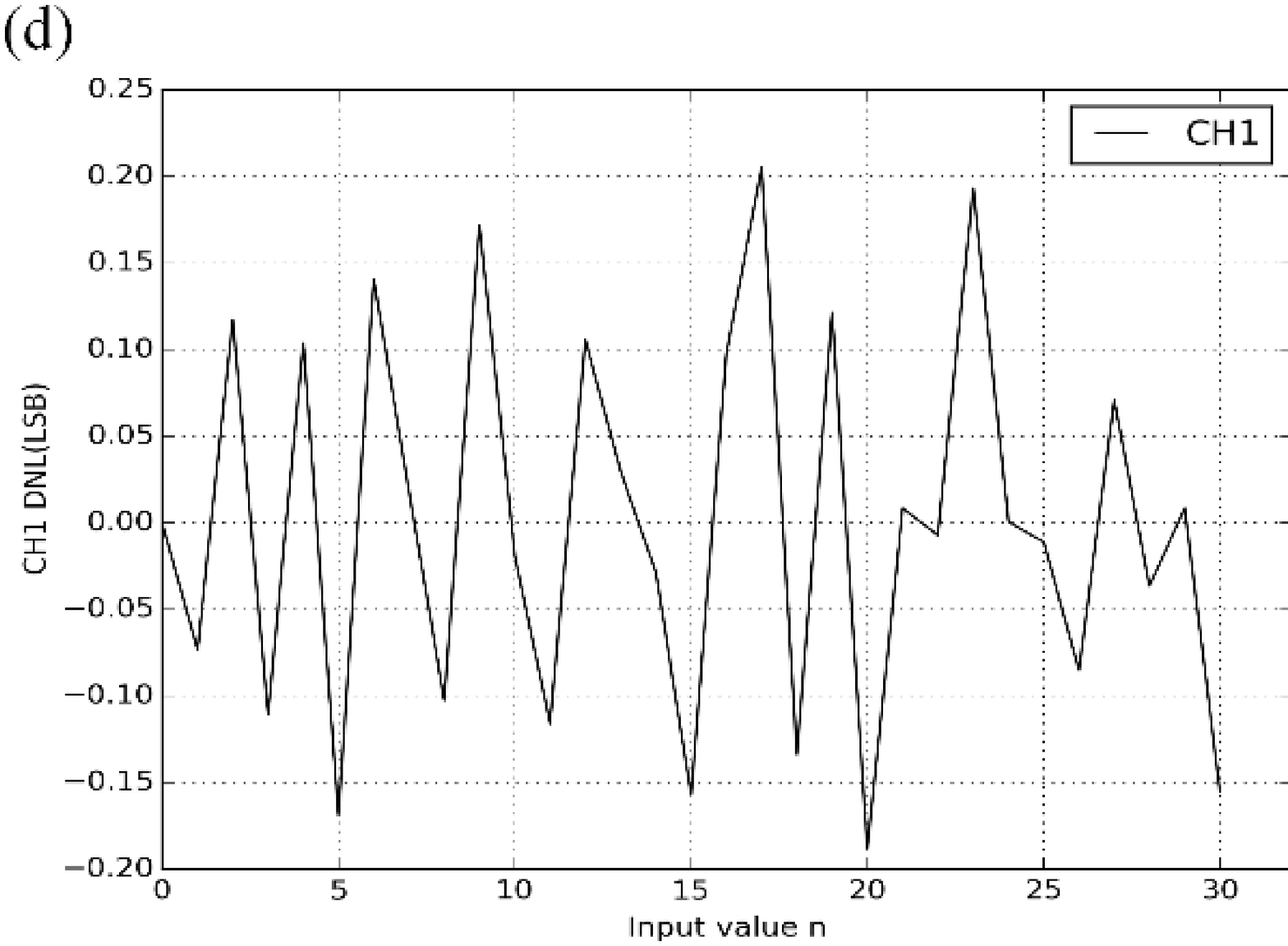}}
  \subfigure{
    \label{Fig.8e}
    \includegraphics[width=3.0in]{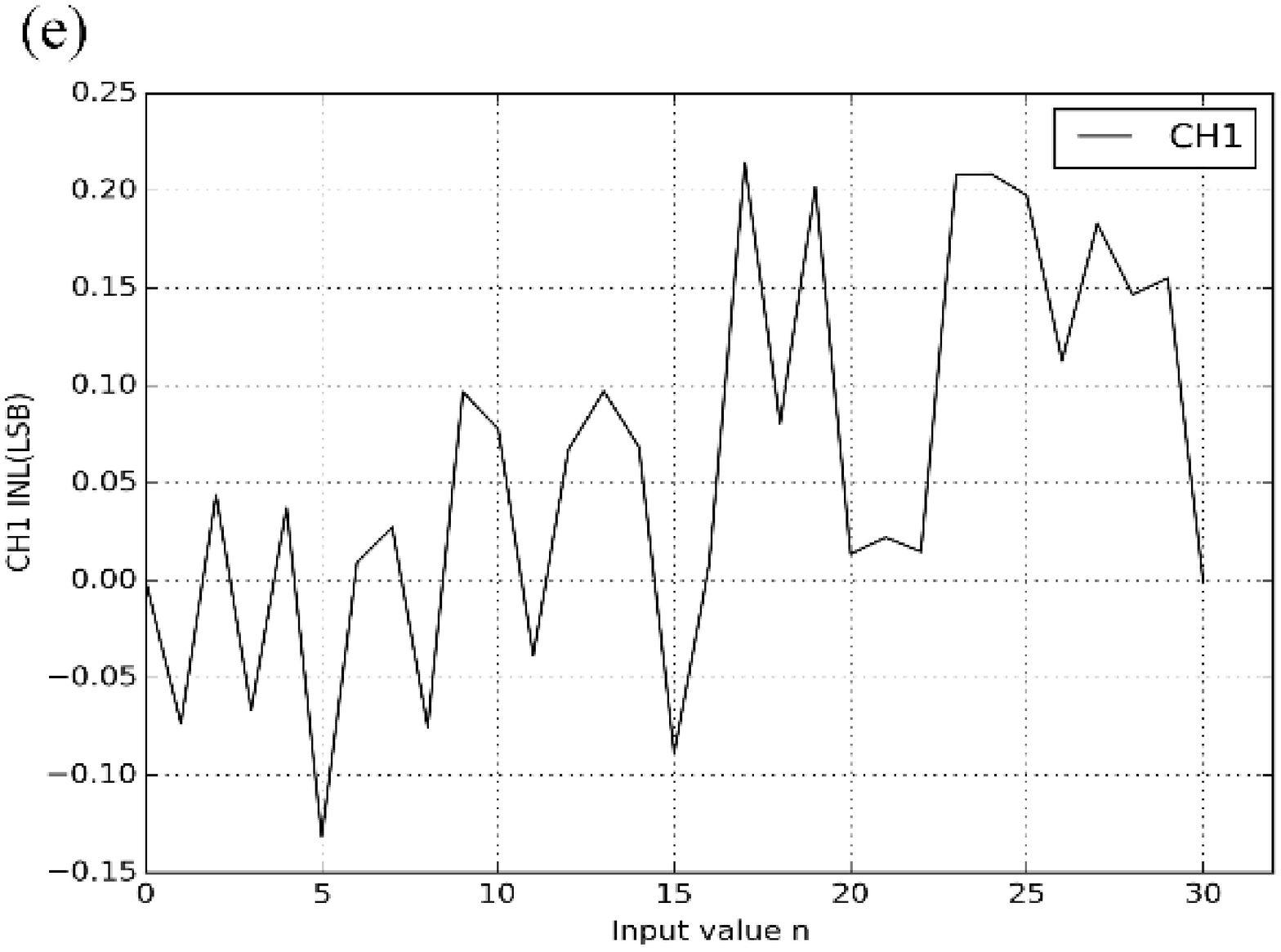}}
  \caption{ Performance of the pulse generator. (a) The oscillogram of the pulse signals from two pulse channels. The dashed red lines are the pulse signal from channel one, and the blue lines is the signal from channel two. (b) The pulse signals with a 50 ps sweeping step. (c) The fine delay time of the delay chain in the pulse generator. (d) The DNL error. (e) The INL error.}
  \label{Fig.8}
\end{figure}

Figure \ref{Fig.8a} shows the oscillogram of the pulse signals from two pulse channels, and the details of the enlarged scale for the signal edges demonstrate the characteristic of the pulses. The figure contains two pairs of pulses: The first pair shows the two channels output pulses with a 50 ps difference in pulse width, and the signal from channel one has a 50 ps delay comparing to channel two; The second pair shows the swap of the pulse width and delay between the two channels. The fine delay of the pulses can be real-time adjusted, thus the pulse generator can be applied to continuously outputting pulse signals in a 50 ps resolution with non-dead-time. The Least-Significant-Bit (LSB) of the pulse generator is measured to be 50 ps. Figure \ref{Fig.8b} shows the pulse signal output from channel one, and the pulse width shows a 50 ps incremental sweeping step. Figure \ref{Fig.8c} shows the exact delay time of the delay chain from the pulse channel one, against the fine bin code. Figure \ref{Fig.8d} and \ref{Fig.8e} show the non-linearity error. The Differential-non-Linearity (DNL) is within -0.19/+0.20 LSB, and the Integrated-non-Linearity (INL) is within -0.13/0.22 LSB. The non-linearity error is small enough and every single bin of the pulse generator is valid.

The time jitter of the pulse generator was also measured. Two pulse channels output two independent pulses and there was a fixed time delay between the two pulses, and the time interval was measured by the TDCs in the device. The main contributors of the Stand-Deviation (STD) of the measured the time intervals include the quantization error of the TDCs, the jitter of both the pulse signals and the jitter of the TDC operating clock. The maximum jitter of the pulse signals should be no more than the STD value. Two independent clocks were separately fed to the pulse generators and the TDCs when measuring the time intervals, thus there was no connection between the pulse signals and the TDC clock. Figure \ref{Fig.9a} shows the STD of the time interval measurement in a wide time interval ranging from 1 ns to 500 ms. The STD goes up when the time interval exceeds 100 ms, due to the limitation of the long term stability of the oscillators. The STD is better than 25 ps in a wide time range of 0~500 ms, and the signal jitter is less than half of the finest 50 ps step. Thus the pulse generator shows a large dynamic range and an extreme stability with a jitter less than 25 ps within a range from 1ns\~{}500 ms.

\subsection{Time-to-digital convertor}

\begin{figure}[h]
  \centering
  \subfigure{
    \label{Fig.9a}
    \includegraphics[width=2.8in]{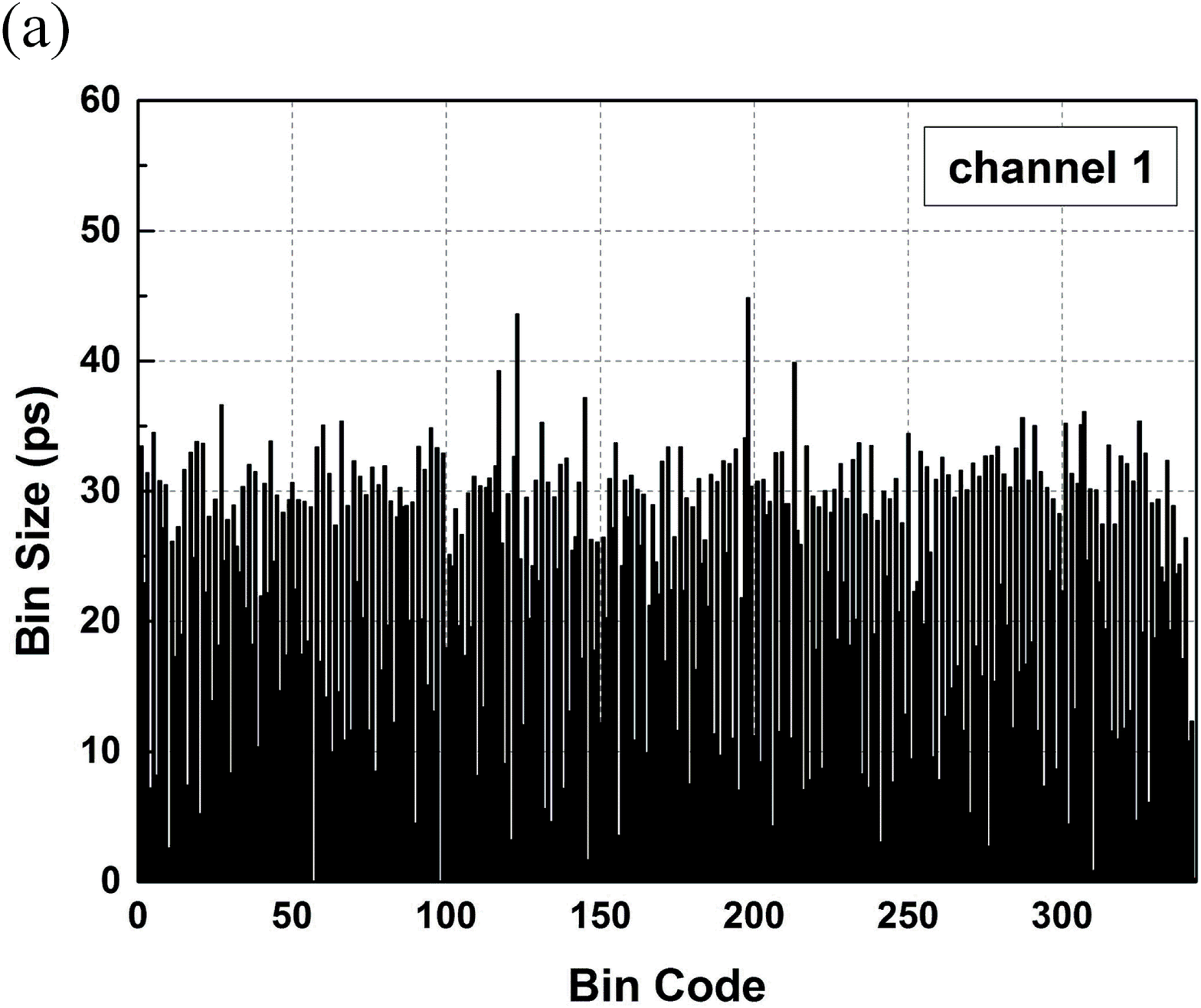}}
  \subfigure{
    \label{Fig.9b}
    \includegraphics[width=2.8in]{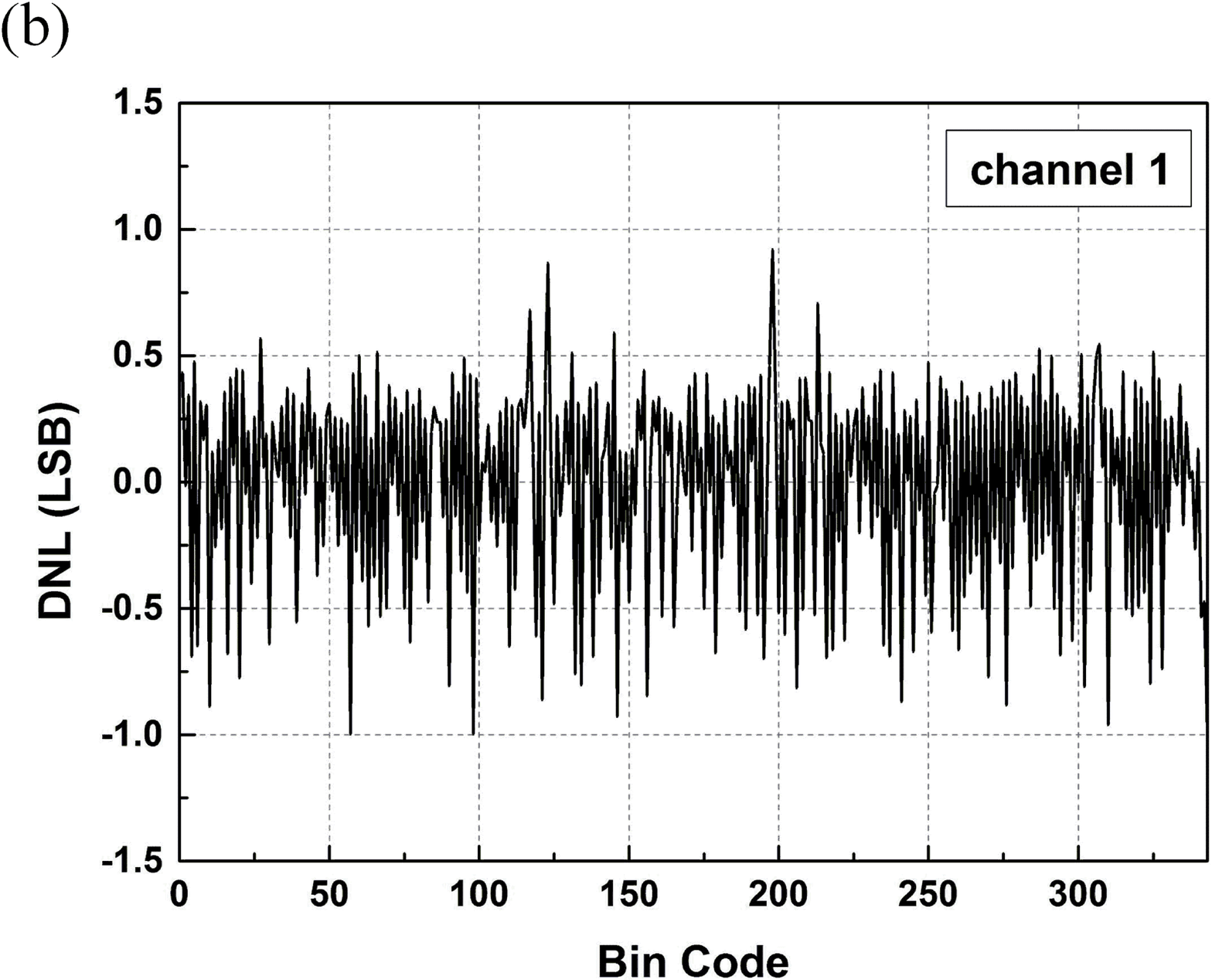}}
  \subfigure{
    \label{Fig.9c}
    \includegraphics[width=2.8in]{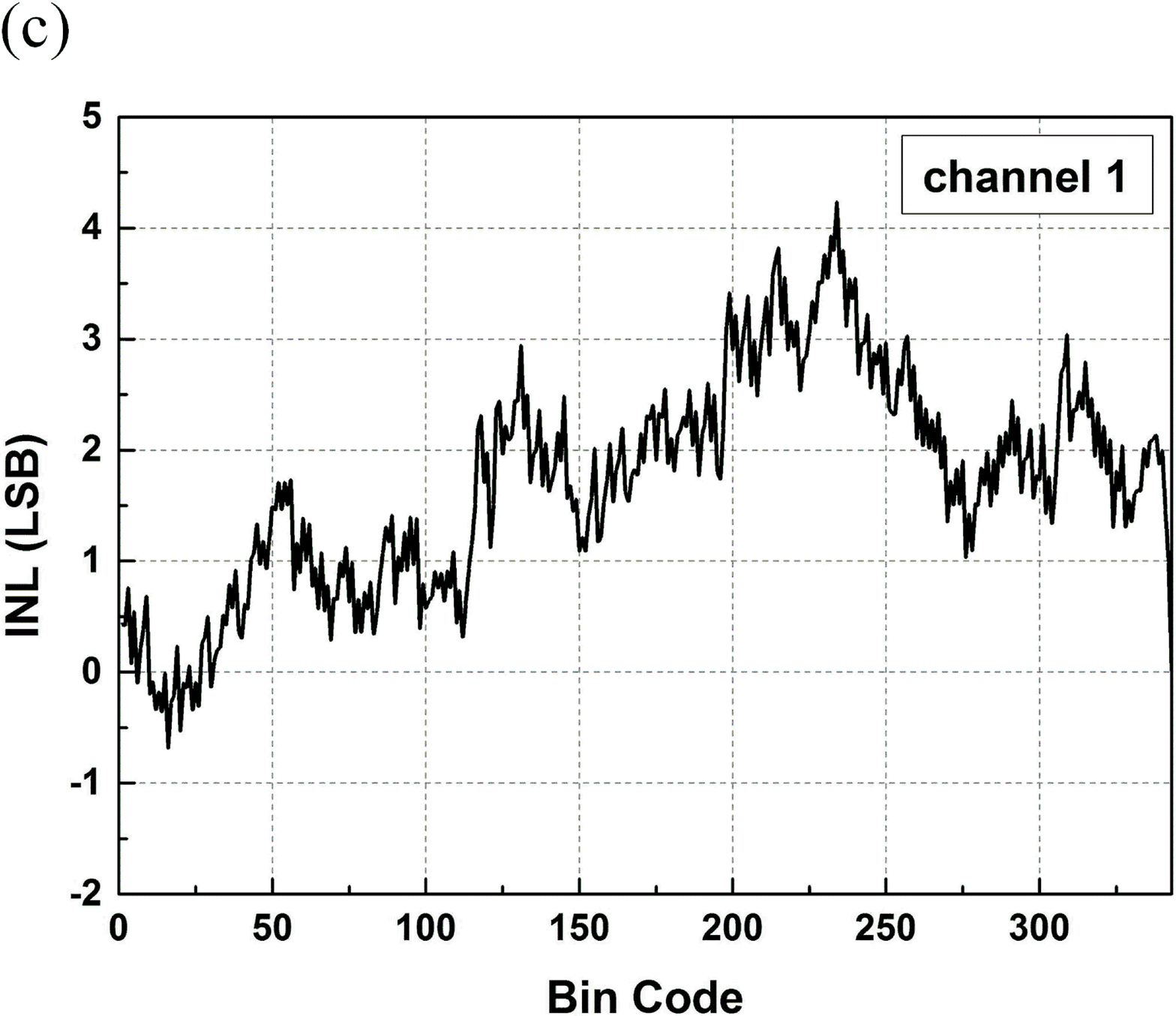}}
  \subfigure{
    \label{Fig.9d}
    \includegraphics[width=2.8in]{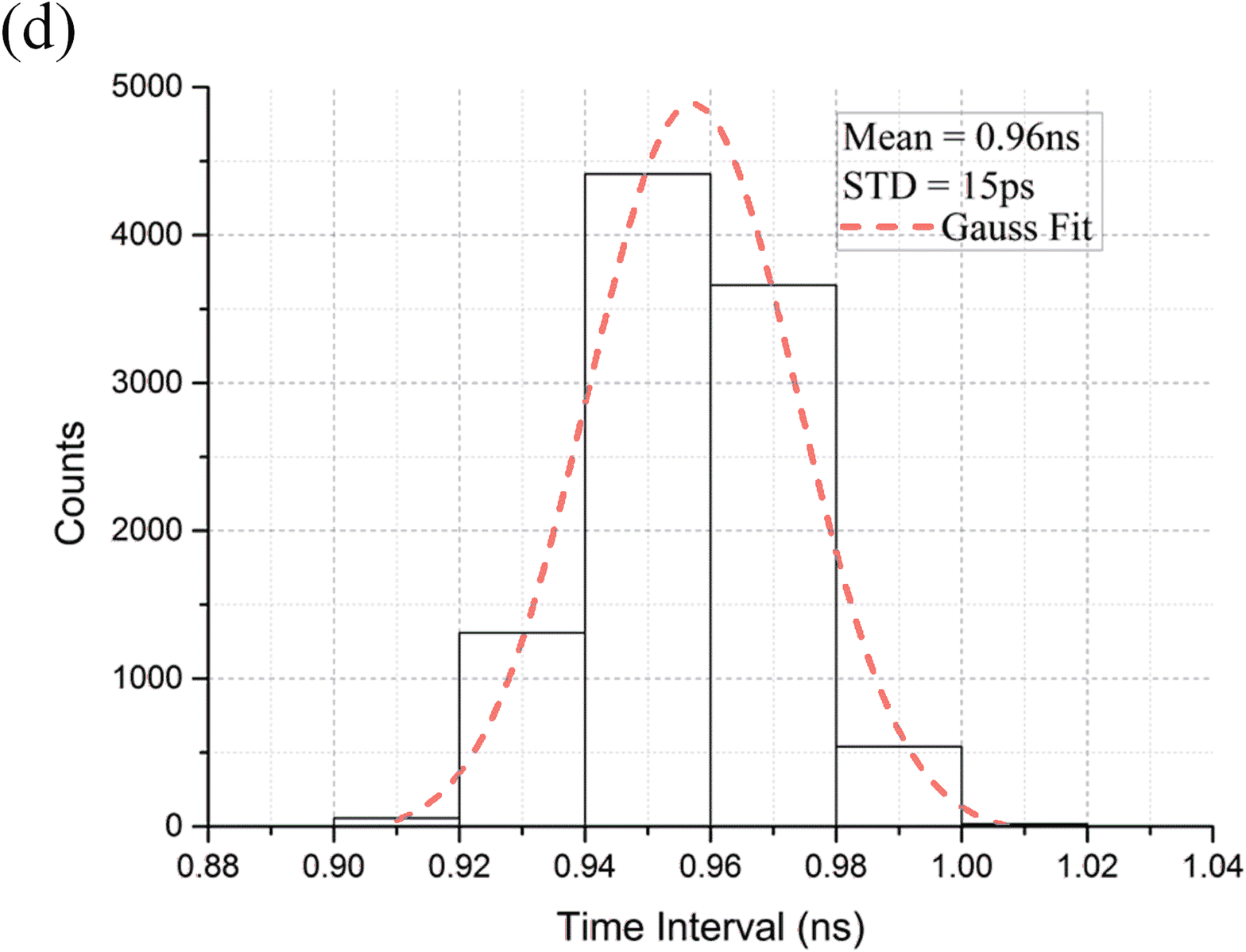}}
  \subfigure{
    \label{Fig.9e}
    \includegraphics[width=2.8in]{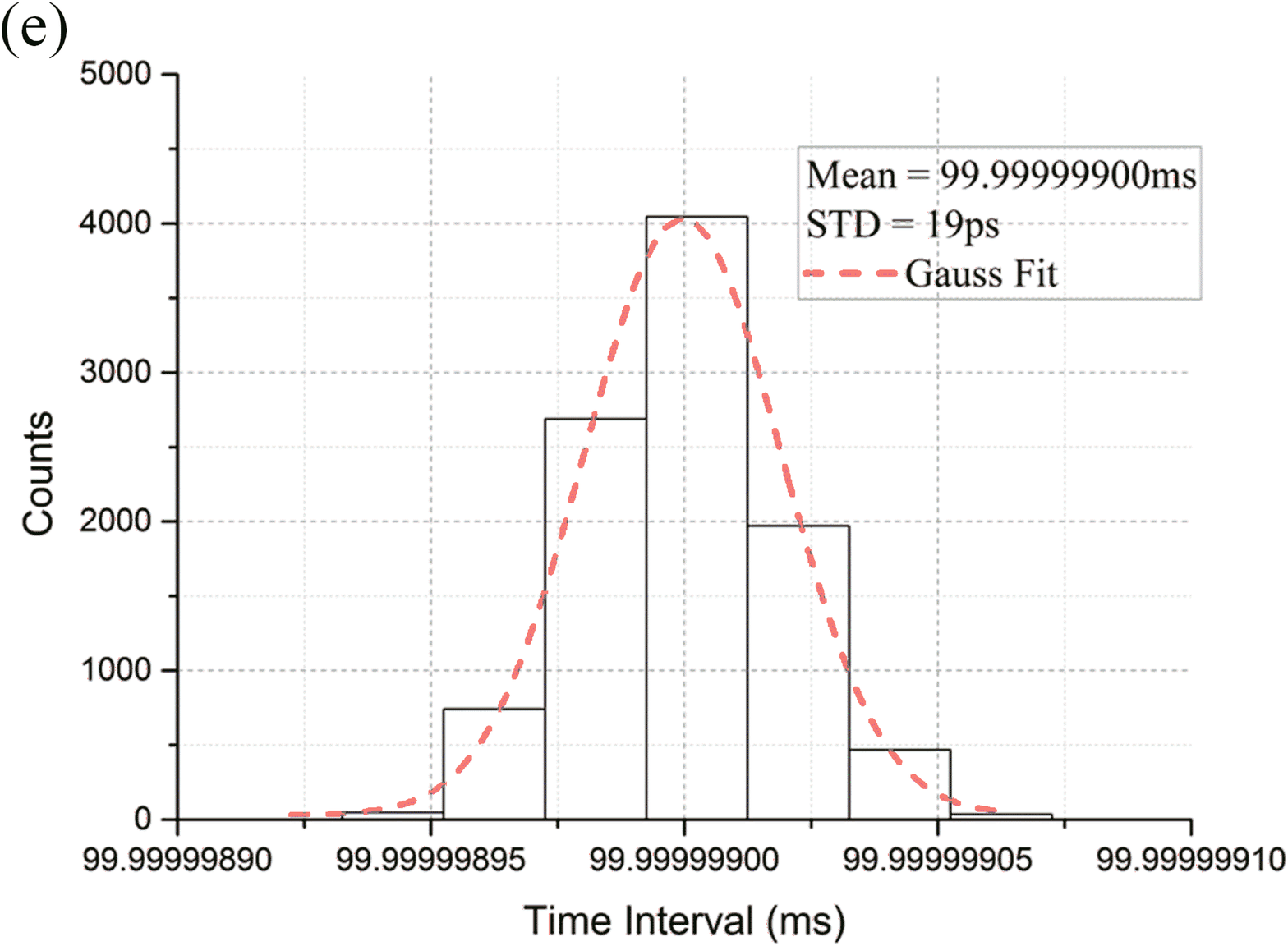}}
  \subfigure{
    \label{Fig.9f}
    \includegraphics[width=2.8in]{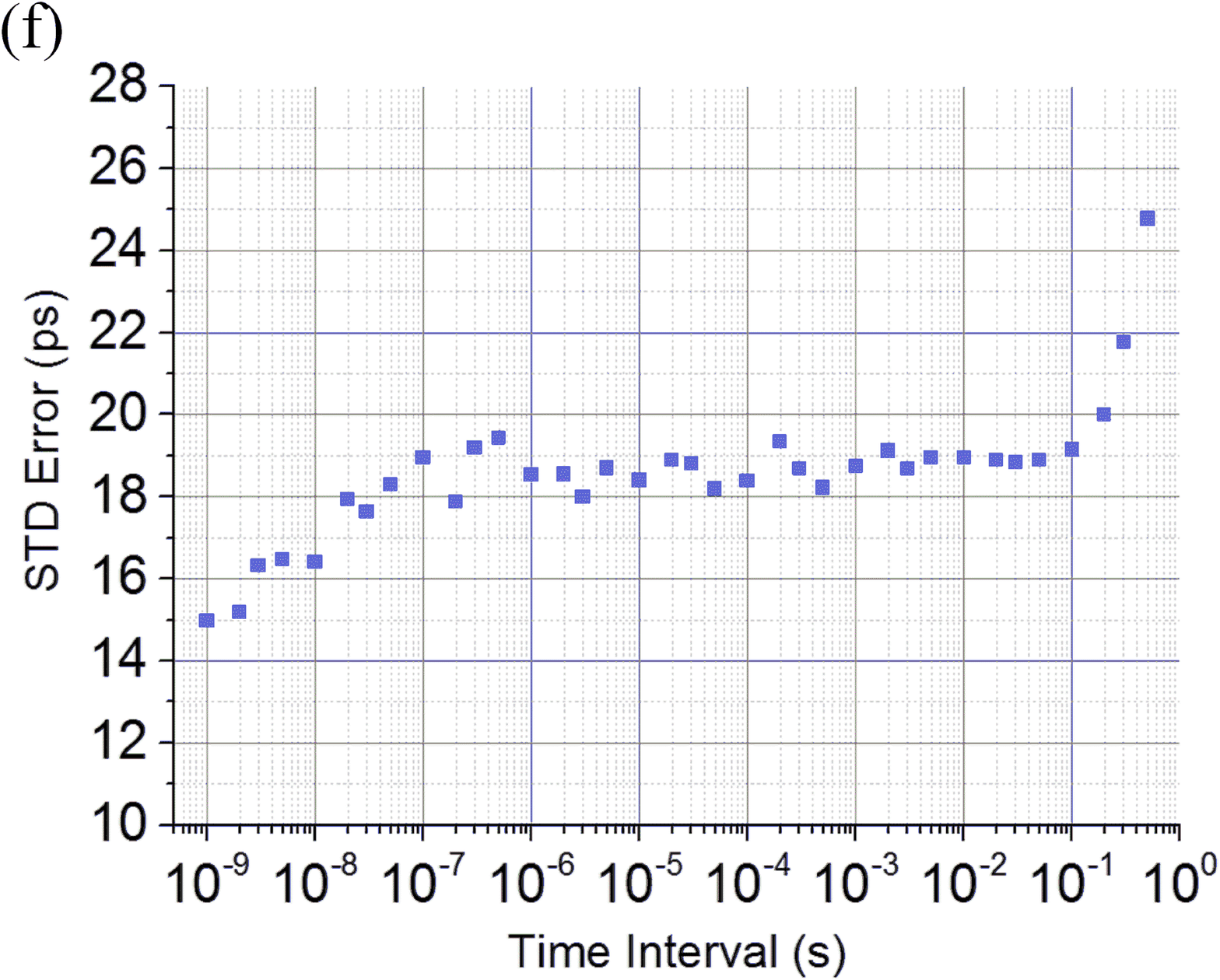}}
  \caption{Characteristic of the FPGA-based time-to-digital convertors. (a) Bin size information of TDC channel 1. (b) The DNL error. (c) The INL error. (d) Statistical histogram when measuring a time interval of 0.96 ns. (e) Statistical histogram when measuring a time interval of 99.999999 ms. (f) The STD error of the time interval measurement in a wide range from 1 ns to 500 ms.}
  \label{Fig.9}
\end{figure}

The bin size and the non-linearity of the TDC were calibrated using the `code-density' method \cite{shubin2010lut}, and the performance was demonstrated through measuring the time interval of the signals input to two TDC channels \cite{szplet2000interpolating}. The arrival time of more than 10¬6 random hits were recorded by one TDC channel in the `code-density' test. The distribution of the measured time data was plotted, and the distribution indicates the bin size information of the channel. Figure \ref{Fig.9a} shows the results of the `code-density' test for channel one of the TDC, and Figure \ref{Fig.9b} and \ref{Fig.9c} show the DNL and INL error, which were calculated using the bin size data. The LSB of the TDC channel corresponds to about 23 ps, with a DNL of -0.95/+0.9 LSB, and an INL of -0.7/+4.2 LSB. A Look-Up-Table (LUT) was generated according to the INL information, and the non-linearity error of time data output from the TDC channel can be corrected using the LUT \cite{shubin2010lut}. Figure \ref{Fig.9d} and \ref{Fig.9e} show the histogram of the time interval measurement. There was a 15 ps STD when measuring an averaged time interval of 0.96 ns, and a 19 ps STD for measuring the 99.999999 ms interval. The input signals were generated by the pulse generator in the device. The STD of the measured time interval not only reflects the pulse jitter, but also proves a high performance TDC. Figure \ref{Fig.9f} shows that the TDC resolution is better than 25 ps in a range from 1ns to 500 ms.

\section{CONCLUSIONS}

We designed and implemented an integrated device with multi-functional generators and TDCs, which can be used in the N-V center based quantum applications including quantum computation and quantum metrology. This device can also be used for other quantum solid state systems, such as quantum dots, phosphorus doped in silicon and defect spins in silicon carbide \cite{claudon2010highly,shields2007semiconductor,gazzano2013bright,castelletto2014n,koehl2011room}. The AWGs in the device are equipped with a 1GB DDR3 memory, and each AWG channel can operate with a 1 Gsps sampling rate and a 16-bit amplitude resolution. The pulse generators are characterized by the superiority of non-dead-time output with a 50 ps high time resolution. The TDCs are designed with a 23 ps time resolution, and an accumulation module is also available. A PC-software which is written in Python is designed to handle the device. We also investigated the performance of the device. The analog outputs from the AWG are free from digital noise with a suppression of better than -143 dBm/Hz. The pulse generator has the desirable performance in non-linearity and jitter, the output pulses show a 50 ps fine step and high stability in a long time range from 1 ns to 500 ms. The TDC also shows quite good time resolution within a large dynamic range up-to 500 ms.
The integrated device shows the ability to achieve a “close-loop” system. The short latency and the good synchronization are great advantages in realizing real-time digital feedback control of quantum systems in the future, just by programming the unused FPGA resources without any hardware modification. The whole device is re-configurable and modularized, thus it can be easily upgraded by replacing the hardware with compatible packages, and re-programming the device. There are plenty of FPGA resources remaining after the implementation, thus the device provides a high feasibility to discover more potential applications. On the other hand, there are bigger Virtex-7 FPGA chips which has more logic resources, and these chips can be assembled in the same pin compatible package, thus the device can be equipped with a more powerful FPGA chip if required in future.

\begin{acknowledgments}
This work was supported by the National Key Basic Research Program of China (Grant No. 2013CB921800 and 2016YFB0501603), the National Natural Science Foundation of China (No. 11275183), the Strategic Priority Research Program (B) of the CAS (Grant No. XDB01030400), and Key Research Program of Frontier Sciences, CAS (Grant No. QYZDY-SSW-SLH004 and QYZDB-SSW-SLH005). X.R. thanks the Youth Innovation Promotion Association of Chinese Academy of Sciences for the support. X.Q. thanks the China Postdoctoral Science Foundation (No. 153727) for the support.
\end{acknowledgments}

%\begin{references}
%\nocite{*}
%\bibliographystyle{unsrt}
\bibliography{bibtex}% Produces the bibliography via BibTeX.
%\end{references}

\end{document}